\documentclass[aps,prc,preprint,groupedaddress]{revtex4}
\usepackage{graphicx}
\usepackage{amsmath}

\begin{document}
\title{Quadrupole and monopole transition properties of $0^+_2$ in Gd isotopes}

\author{Masayuki Matsuzaki}
\email[]{matsuza@fukuoka-edu.ac.jp}
\affiliation{Department of Physics, Fukuoka University of Education, 
             Munakata, Fukuoka 811-4192, Japan}
\author{Tomoya Ueno}
\affiliation{Department of Physics, Fukuoka University of Education, 
             Munakata, Fukuoka 811-4192, Japan}

\date{\today}

\begin{abstract}
The longstanding problem of characterization of the $0^+_2$ states in Gd isotopes is revisited 
by adopting the Nilsson$+$BCS mean field and the random-phase approximation. The interband 
electric quadrupole transition strengths varying almost two orders of magnitude are nicely 
reproduced at the same time as other observables. 
These results indicate that the $0^+_2$ states, in particular, those in lighter isotopes 
are well described as $\beta$ vibrations excited on top of deformed ground states without 
recourse to the shape-coexistence picture.

Subject index D12, D13
\end{abstract}

\maketitle

\section{Introduction}

The first $0^+$ excitation, denoted as $0^+_2$, is one of the fundamental excitations in 
atomic nuclei. It carries information about the nuclear shape and the pairing correlation. 
In medium and heavy nuclei, ground states of those which are off closed shell are more or 
less deformed. In the traditional picture of Bohr and Mottelson, their deformations are 
axially symmetric and the $\beta$ ($K^\pi=0^+$) and $\gamma$ ($K^\pi=2^+$) vibrations exist as 
low-lying collective excitations~\cite{BM}. Actually the latter has been widely confirmed in 
the nuclear chart. In contrast, properties of observed $0^+_2$ are still controversial~\cite{Ga}. 
The most decisive observable is $B(E2,0^+_2\rightarrow2^+_1)$. But its data often have 
relatively large error bars and they vary strongly with nucleon numbers. 

Not only as a particle-hole collective state, the shape vibration, but $0^+$ states can 
be excited as a particle-particle collective state, pairing vibration, via two-nucleon 
(2n) transfer reactions. In superfluid nuclei, their typical crosssection to $0^+_2$ is 
estimated as $2\%$ of that to $0^+_1$~\cite{BB}. In the 70's, a lot of $(p,t)$ and $(t,p)$ 
experiments were done and found that transfer crosssections to $0^+_2$ became comparable 
with those to $0^+_1$ in transitional $N=88 - 90$ nuclei, Ref.~\cite{FGH} for example of Gd. 
Their results were interpreted mainly in terms of the shape-coexistence picture~\cite{HW,LNV}, 
for example. On the other hand, a large-amplitude shape fluctuation encompassing two minima, 
if exist, was also conjectured~\cite{DH}. 

Another observable that is known to be sensitive to shape deformation and coexistence is 
$\rho^2(E0)$, the reduced electric monopole transition strength measured through internal 
electron conversions~\cite{Ra,WZC}. The latter reference discussed a wide variety of 
medium and heavy nuclei based on available data of $0^+\rightarrow0^+$, $2^+\rightarrow2^+$, 
and $4^+\rightarrow4^+$ transitions. On the other hand, the $E0$ transition strength is one of 
indicators of the cluster structure in light nuclei~\cite{YFH}. Reference~\cite{KS} 
compiles $0^+\rightarrow0^+$ transition data throughout the nuclear chart. Theoretically 
the $E0$ strengths in medium-heavy nuclei have been systematically studied mainly 
in terms of the interacting boson approximation (IBA) model, \cite{BWC}, for example. 
But characterization of $0^+_2$ is still not decisive. This suggests that not only 
properties of $0^+_2$, such as the level energy, $E2$ and $E0$ transitions to the ground band, 
but also other information such as rotational band structure should be taken into account. 
Studies aiming at such a direction were pursued for $^{152}$Sm, for example, 
\cite{CWR,Bu,CCD}. In addition, the relation between the properties of the $0^+_2$ in $^{154}$Gd and 
the spectra of an adjacent odd-$A$ nucleus was also argued trying to discriminate different 
pictures~\cite{SS1,SS2}. 

In the following, we study $B(E2,0^+_2\rightarrow2^+_1)$, $\rho^2(E0,0^+_2\rightarrow0^+_1)$, 
and their ratio, $X(E0/E2)$ of Gd isotopes that is one of the isotope chains 
about which the richest 
information is available, paying attention also to rotational properties. Experimental 
data are taken from the Live Nuclear Chart of IAEA~\cite{IAEA} for level energies and 
$B(E2)$, and from Ref.~\cite{KS} for $\rho^2(E0)$. 

\section{The model}

We adopt a traditional mean field $+$ random-phase approximation (RPA). The mean field 
is the Nilsson $+$ BCS model, 
\begin{gather*}
h=h_\mathrm{Nil}-\Delta_\tau (P_\tau^\dagger+P_\tau)
                   -\lambda_\tau N_\tau, \quad \tau\in\{n,p\}, \\
h_\mathrm{Nil}=\frac{\mathbf{p}^2}{2M}
                +\frac{1}{2}M(\omega_x^2 x^2 + \omega_y^2 y^2 + \omega_z^2 z^2)
                +v_{ls} \mathbf{l\cdot s} 
                +v_{ll} (\mathbf{l}^2 - \langle\mathbf{l}^2\rangle_{N_\mathrm{osc}}),
\end{gather*}
where standard notations for each quantity are understood. The $\mathbf{l\cdot s}$ and 
$\mathbf{l}^2$ terms are given by the singly-stretched coordinates. Their strengths are 
taken from Ref.~\cite{BR}. The deformation of the oscillator potential is parameterized as 
\begin{gather*}
\omega_j=\omega_0\Large[
1-\frac{2}{3}\epsilon_2\cos{\large(\gamma+\frac{2\pi\nu_j}{3}\large)}
\Large], \quad j\in\{x,y,z\}, \\
\nu_x=1, \quad \nu_y=-1, \quad \nu_z=0,
\end{gather*}
where $\omega_0$ is determined so as to conserve the nuclear volume. 
The cranking term $-\hbar\omega_\mathrm{rot}J_x$ is also introduced when necessary. 
The residual pairing (P) plus isoscalar doubly-stretched quadrupole-quadrupole (Q-Q) 
interaction is given by 
\begin{equation*}
H_\mathrm{int}=-G_\tau\widetilde{P_\tau}^\dagger \widetilde{P_\tau}
               -\frac{1}{2}\sum_{K=0}^2 \kappa_K^{(+)}\widetilde{Q_K''^{(+)}}^\dagger \widetilde{Q_K''^{(+)}},
\end{equation*}
where $Q_K''^{(+)}$ are obtained from the spherical harmonics as 
\begin{gather*}
Q_{2\mu}(\mathbf{r})=r^2Y_{2\mu}(\theta,\phi), \\
Q_K^{(+)}(\mathbf{r})=\frac{1}{\sqrt{2(1+\delta_{K0})}}\left(Q_{2K}(\mathbf{r})+Q_{2-K}(\mathbf{r})\right), \\
Q_K''^{(+)}=Q_K^{(+)}\left(x_j\rightarrow x_j'' = \frac{\omega_j}{\omega_0}x_j\right).
\end{gather*}
Here '$\,\widetilde{\quad}\,$' indicates that the ground-state expectation values 
are subtracted, and the $K\neq0$ terms mix into the mode under consideration when 
a rotation and/or $\gamma$ deformation are introduced. The original P$+$Q-Q interaction 
was used to determine the parameters of Bohr's collective Hamiltonian~\cite{KB} and 
applied to deformed Gd isotopes in Ref.~\cite{GKH}. The doubly-stretched Q-Q interaction 
was proposed to fulfill a shape selfconsistency in deformed nuclei and was shown 
to be effective in actual description of deformed nuclei~\cite{KMY,SK}, and 
further extended to rotating nuclei~\cite{SM1}. 

Transition strengths are calculated as follows. 
For the initial state $|\mathrm{i}\rangle=|0^+_2\rangle=X^\dagger|0^+_1\rangle$ 
and the final state $|\mathrm{f}\rangle=|2^+_1\rangle$, where $2^+_1$ is the first excited 
member of the ground-state band, the $E2$ transition strength is given by 
\begin{equation*}
B(E2,I_\mathrm{i}K_\mathrm{i}=0\rightarrow I_\mathrm{f}K_\mathrm{f}=0)
=\langle I_\mathrm{i}020|I_\mathrm{f}0\rangle^2|\langle[Q_0^{(+)},X^\dagger]\rangle_\mathrm{RPA}|^2,
\end{equation*}
when the rotational effect, the difference between the intrinsic states of $|0^+_1\rangle$ 
and $|2^+_1\rangle$, is ignored. Here $\langle I_\mathrm{i}020|I_\mathrm{f}0\rangle=1$ for 
$I_\mathrm{i}=0$, $I_\mathrm{f}=2$, and $\langle[\cdot,X^\dagger]\rangle_\mathrm{RPA}$ denotes 
the transition amplitude associated with the RPA phonon $X^\dagger$. The rotational effect is 
taken into account by the method~\cite{SN} based on the generalized intensity relation 
(GIR)~\cite{BM}. The GIR indicates that the angular-momentum dependence of interband transition 
matrix elements fits into the form, 
$M_1+M_2[I_\mathrm{f}(I_\mathrm{f}+1)-I_\mathrm{i}(I_\mathrm{i}+1)]$. 
In the textbook of Bohr and Mottelson, $M_1$ and $M_2$ are obtained by fitting to the data. 
Such a fitting was also done for adjacent nuclei recently~\cite{CCD}. 
Reference~\cite{SN} proposed a method to represent it in terms of intrinsic matrix elements 
given by the mean field and RPA. The concrete form for the present case is given by replacing 
\begin{align*}
M_1&=\langle[Q_0^{(+)},X^\dagger]\rangle_\mathrm{RPA}\rightarrow \\
M_1&+M_2[I_\mathrm{f}(I_\mathrm{f}+1)-I_\mathrm{i}(I_\mathrm{i}+1)] \\
&=\langle[Q_0^{(+)},X^\dagger]\rangle_\mathrm{RPA}
+\frac{\hbar}{2\sqrt{3}\mathcal{J}}
  \frac{\mathrm{d}}{\mathrm{d}\omega_\mathrm{rot}}\langle[Q_1^{(+)},X^\dagger]\rangle_\mathrm{RPA}
[I_\mathrm{f}(I_\mathrm{f}+1)-I_\mathrm{i}(I_\mathrm{i}+1)],
\end{align*}
where $\mathcal{J}$ is the moment of inertia of the ground-state band. 

The non-dimensionalized $E0$ transition matrix element from 
$|\mathrm{i}\rangle=|0^+_2\rangle=X^\dagger|0^+_1\rangle$ 
to $|\mathrm{f}\rangle=|0^+_1\rangle$ is given by 
\begin{equation*}
\rho(E0,\mathrm{i}\rightarrow\mathrm{f})
=\langle[r^2,X^\dagger]\rangle_\mathrm{RPA}/eR^2,
\end{equation*}
with $R=r_0A^{1/3}$. Effective charges are not introduced, and $Q_K^{(+)}$ and $r^2$ in 
these expressions are understood as their proton part multiplied by $e$. 
The $X$ ratio~\cite{Ra} is defined by 
\begin{equation*}
X(E0/E2)=\frac{(\rho(E0))^2e^2R^4}{B(E2)}.
\end{equation*}
\section{Results and discussion}

\subsection{The shape of ground states}

First of all, quadrupole deformations of the ground states of $^{146-160}_{82-96}$Gd 
are determined by using the Nilsson-BCS-Strutinsky method~\cite{BRA} assuming $\epsilon_4=0$, 
where $\epsilon_4$ is the magnitude of the hexadecapole deformation of the mean field. 
Calculations were done adopting five major shells $N_\mathrm{osc}=$ 4 -- 8 
for neutrons and $N_\mathrm{osc}=$ 3 -- 7 for protons. Obtained $\epsilon_2$ are 
summarized in TABLE~\ref{table1}. For all the cases, $\gamma=0$. 

\begin{table}[htbp]
 \caption {The deformation of the ground state determined by the Nilsson-BCS-Strutinsky 
method and the experimental data used to determine the properties of Gd isotopes. 
Among them, the pairing gaps are obtained by the third difference of the experimental masses. }
 \label{table1}
\begin{center}
\begin{tabular}{ccccccc} \hline\hline
$N$ & $\epsilon_2$ & $\Delta_\mathrm{n}$ (MeV) & $\Delta_\mathrm{p}$ (MeV) & $E_{0^+_2}$ (MeV) & $E_{2^+_\gamma}$ (MeV) & $E_{2^+_1}$ (MeV) \\ 
\hline
86 & $\quad$0.07 & 1.00 & 1.42 & 1.207 & 1.430 & 0.638 \\
88 & $\quad$0.18 & 1.11 & 1.48 & 0.615 & 1.109 & 0.344 \\
90 & $\quad$0.22 & 1.28 & 1.13 & 0.681 & 0.996 & 0.123 \\
92 & $\quad$0.25 & 1.07 & 0.96 & 1.049 & 1.154 & 0.089 \\
94 & $\quad$0.27 & 0.89 & 0.88 & 1.196 & 1.187 & 0.080 \\
96 & $\quad$0.26 & 0.83 & 0.85 & 1.380 & 0.988 & 0.075 \\
\hline\hline
\end{tabular}
\end{center}
\end{table}

The lightest two, $^{146}$Gd and $^{148}$Gd, are spherical, $\epsilon_2=0$, as expected 
and they are omitted in TABLE~\ref{table1} and the following calculation. 
Next one, $^{150}$Gd, is almost spherical, $\epsilon_2=0.07$. 
Experimentally the two-phonon triplet in terms of 
spectra of spherical nuclei splits to some extent and the $0^+$ among them was 
labeled as the quasi-$\beta$~\cite{HS}. A boson expansion calculation included in 
that reference shows more developed rotational character. Then we included $^{150}$Gd 
in the following figures for the sake of comparison but obviously the $N$ dependence 
is discontinuous to $^{152}$Gd and heavier. 

In the literature, the ground state of $^{152}$Gd has been said to be spherical and 
this lead to the shape-coexistence interpretation of the 2n-transfer data~\cite{FGH}, 
but the present result of the Strutinsky method disagrees. In the following we put 
some emphasis on this issue. 

\subsection{Transition strengths}

\subsubsection{$N$ dependence}

Adopting the deformation $\epsilon_2$ obtained above, the mean field plus RPA calculations 
are performed in three major shells $N_\mathrm{osc}=$ 4 -- 6 for neutrons and 
$N_\mathrm{osc}=$ 3 -- 5 for protons, which give phenomenologically appropriate results, 
as in Ref.~\cite{SN}. Interaction strengths $G_\mathrm{n}$, $G_\mathrm{p}$, and $\kappa_K^{(+)}$ 
($K=0,2$) are adjusted to reproduce experimental pairing gaps, $E_{0^+_2}$, and $E_{2^+_\gamma}$ 
tabulated in TABLE~\ref{table1}. The $K=1$ component is adjusted to give zero energy 
to the Nambu-Goldstone mode. 

\begin{figure}[htbp]
 \includegraphics[width=7cm]{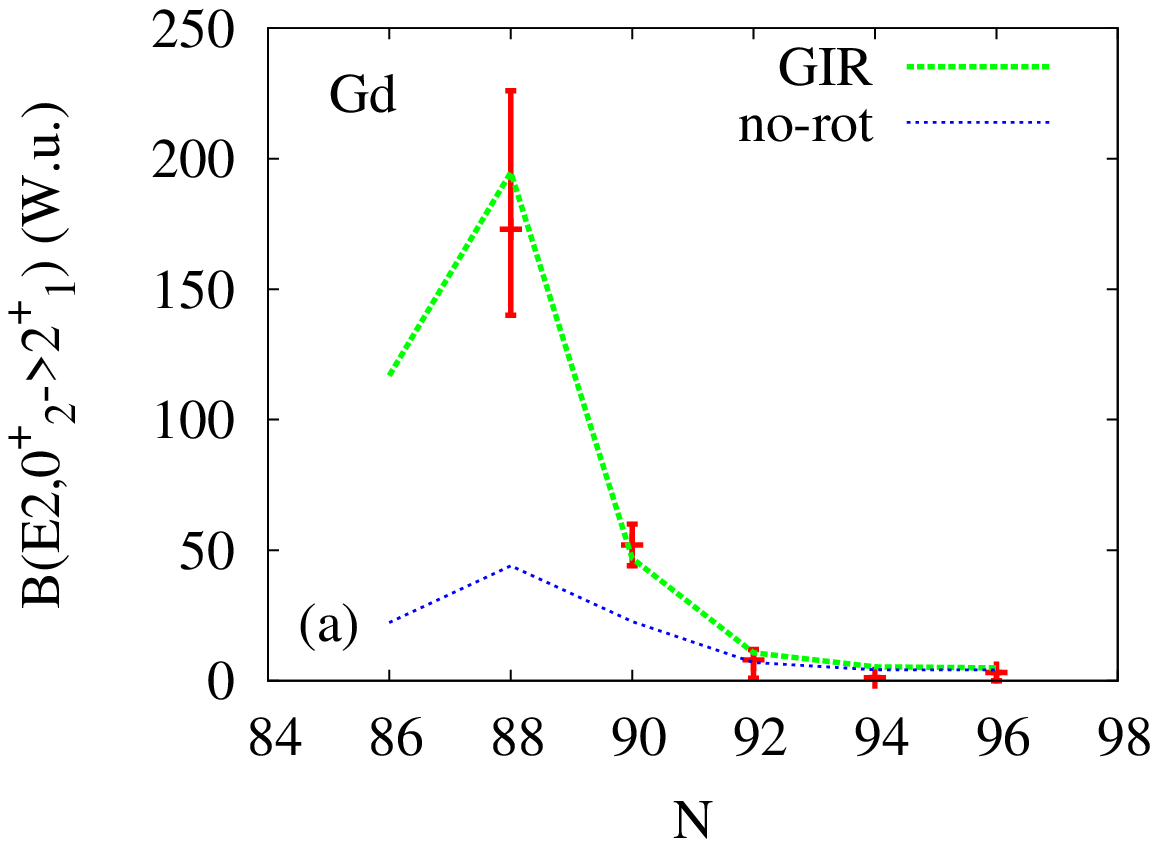}
 \includegraphics[width=7cm]{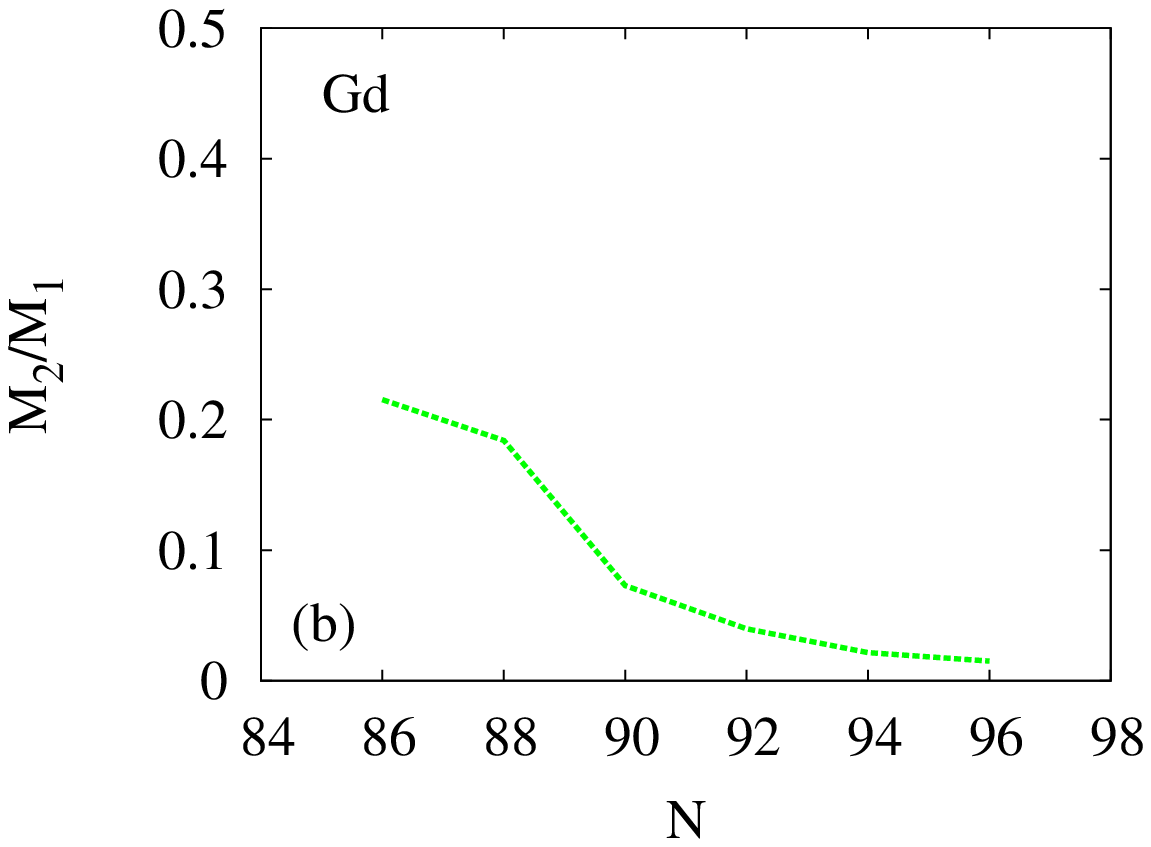}
 \caption{(Color online) (a) Experimental and calculated $B(E2,0^+_2\rightarrow2^+_1)$ in the 
Weisskopf unit as functions of the neutron number of Gd isotopes. Green dashed and blue 
dotted curves represent the calculations with and without the rotational effect given by 
the method based on the GIR. Data are taken from Ref.~\cite{IAEA}. (b) The ratio $M_2/M_1$ 
that gives the magnitude of the rotational effect on the transition matrix element.}
 \label{fig1}
\end{figure}

Figure~\ref{fig1}(a) presents the most important quantity to characterize $0^+_2$, $B(E2)$ 
to the ground band. Results of calculations with and without inclusion of the rotational 
effect are compared with the data. They vary almost two orders of magnitude. This steep 
variation is nicely reproduced by including the rotational effect. Its significance 
is shown in Fig.~\ref{fig1}(b) by the ratio $M_2/M_1$. The effect is conspicuous 
in lighter isotopes because of its dependence on $\mathcal{J}$ determined from 
$E_{2^+_1}=2\cdot(2+1)\hbar^2/2\mathcal{J}$ in TABLE~\ref{table1}. This is very contrastive to 
the $\gamma$ vibration for which the corresponding 
$\frac{5}{2}\times B(E2,2^+_\gamma\rightarrow0^+_1)$ stays within 10 -- 20 W.u. 
(Fig.~\ref{fig5}(b)). From this largeness of $B(E2,0^+_2\rightarrow2^+_1)$, the $0^+_2$ 
states in $^{152}$Gd and $^{154}$Gd have been thought of as typical $\beta$ 
vibrations~\cite{CWR,Ga,KWK1} but a different interpretation was also proposed as 
discussed later. The smallness in heavier isotopes, already presented in Ref.~\cite{SN}, 
will also be discussed later. 

\begin{figure}[htbp]
 \includegraphics[width=7cm]{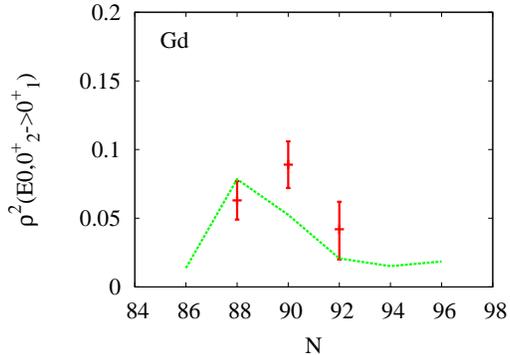}
 \caption{(Color online) Experimental and calculated $\rho^2(E0,0^+_2\rightarrow0^+_1)$ 
as functions of the neutron number of Gd isotopes. The rotational effect does not appear 
in this quantity. Data are taken from Ref.~\cite{KS}.}
 \label{fig2}
\end{figure}

Figure~\ref{fig2} compares the result for $\rho^2(E0,0^+_2\rightarrow0^+_1)$ with the 
available data~\cite{KS}. Note that Refs.~\cite{LSS,Ge} included a data point of 
$\rho^2(E0,0^+_2\rightarrow0^+_1)$ in their calculation for $^{158}$Gd, but this is 
actually that of $\rho^2(E0,2^+_2\rightarrow2^+_1)$. See Refs.~\cite{WZC,GRB}. 
A recent large-scale calculation adopting the constrained Hartree-Fock-Bogoliubov 
theory with the Gogny D1S interaction~\cite{DGL} results in failure to reproduce 
the order of magnitudes of the observed $\rho^2(E0)$. Then we have to have recourse to 
more phenomenological models to discuss their actual isotope dependence. 
In the literature, the IBA model~\cite{ZIB} and the geometrical coherent-state 
model~\cite{RRM} reproduce the data well. The present calculation gives similar results. 

The next aspect is the isotope dependence. Preceding the data for $N=92$, 
Ref.~\cite{BWC} argued that the rise from $N=88$ to 90 as well as in other isotope 
chains is a signal of the spherical-deformed shape phase transition and consequently 
$\rho^2(E0)$ would stay large in heavier isotopes. Unfortunately this has not been proved 
to apply. In the present calculation, the maximum occurs at $N=88$ not 90. 

\begin{figure}[htbp]
 \includegraphics[width=7cm]{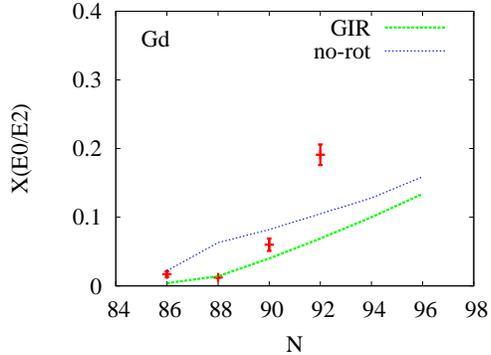}
 \caption{(Color online) Experimental and calculated non-dimensionalized ratio 
of $E0$ and $E2$ transition strengths graphed in the same manner as Fig.~\ref{fig1}(a). 
Data are taken from Ref.~\cite{KS}.}
 \label{fig3}
\end{figure}

Figure~\ref{fig3} compares the calculated $X$ ratios to the data. The reason why 
Refs.~\cite{BGS,BAB} included a data point of $^{158}$Gd is the same as above. 
This figure indicates that the isotope dependence is predominantly determined 
by the denominator. The present calculation reproduces the rising trend but it is 
quantitatively weaker. This comes from the result that $B(E2)$ in heavier isotopes 
looks to be larger than the data. This point will be discussed later. 
The discontinuity between $N=86$ and 88 seen in Figs.~\ref{fig1}(a) and \ref{fig2} 
disappears because both the denominator and numerator vary to a similar extent. 

\subsubsection{Individual nucleus}

(1) $^{152}$Gd

In the literature, ground states of $N=88$ isotones have been considered to be spherical, 
see for example, Ref.~\cite{DH} for Sm and \cite{FGH} for Gd. However, the observed 
in-band $B(E2,2^+_1\rightarrow0^+_1)=73^{+7}_{-6}$ W.u.~\cite{IAEA} suggests a 
moderate deformation and actually in the present calculation, the rotational-model 
expression gives 
\begin{equation*}
B(E2,2^+_1\rightarrow0^+_1)=(eQ_0)^2=75\,\mathrm{W.u.}
\end{equation*}
In this nucleus, high-spin states of the $0^+_1$ and $0^+_2$ bands were studied in the 
2000s~\cite{WHM,CLR}. These works show smooth behavior of these bands starting from 
the bandhead with gradual stretching. Moreover, a $g$ factor measurement of 
$2^+_1$ -- $6^+_1$ also supports rotational character of the low-spin members of 
the ground-state band~\cite{MKH}. 
Actually the present calculation gives a smooth behavior as a function of the 
rotational frequency, for example, 
$g=\frac{\langle\mu_x\rangle}{\langle J_x\rangle}=0.41$ at 
$\hbar\omega_\mathrm{rot}=\frac{E_{2^+_1}}{2}=0.172$ MeV, which is very close 
to the collective value, $g_\mathrm{R}=\frac{Z}{A}$. 

\begin{figure}[htbp]
 \includegraphics[width=7cm]{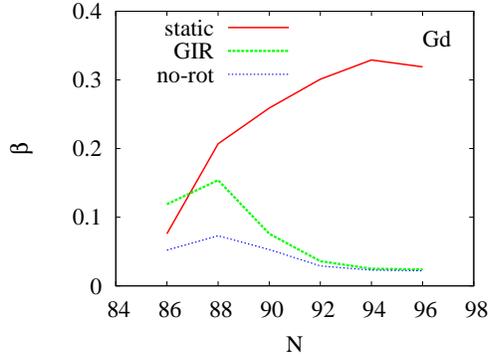}
 \caption{(Color online) Calculated static deformation $\beta_\mathrm{s}$ (red solid), 
zero-point amplitudes $\beta_0$ of the $\beta$ vibration with (green dashed) and without 
(blue dotted) the rotational effect as functions of the neutron number of Gd isotopes.}
 \label{fig4}
\end{figure}

Next, an implication of the conspicuous magnitude of $B(E2,0^+_2\rightarrow2^+_1)$ is 
mentioned. As first discussed by Kumar~\cite{Ku}, not only in-band but also interband 
$B(E2)$ brings information about the deformation. Based on this, the model-independent 
effective deformation, $\beta_\mathrm{eff}$, is examined and compared with the IBA 
model~\cite{WWC}. According to this work, the square of the effective deformation 
of the $0^+_2$ state is given by 
\begin{equation*}
|\beta_\mathrm{eff}|^2=\frac{\sum_jB(E2,0^+_2\rightarrow2^+_j)}{(\frac{3}{4\pi}ZeR^2)^2}.
\end{equation*}
The summation is expected to be almost saturated with $j=1$ and 2. In the present mean field 
plus RPA model, the $j=2$ term gives the static deformation of the $0^+_2$ state while 
the $j=1$ term gives the zero-point amplitude of the $\beta$ vibration. Those converted 
by
\begin{equation*}
\beta_0=\frac{\sqrt{B(E2,0^+_2\rightarrow2^+_1)}}{\frac{3}{4\pi}ZeR^2}
\end{equation*}
from the $B(E2)$ values in Fig.~\ref{fig1}(a) are compared with corresponding static deformation, 
\begin{equation*}
\beta_\mathrm{s}=\frac{\langle Q_0\rangle_\mathrm{IS}}{\frac{3}{4\pi}AR^2}
\end{equation*}
in Fig.~\ref{fig4}. Here the subscript designates the isoscalar quadrupole moment. 

The RPA is a small-amplitude approximation. 
It is not obvious from the ratio of $\beta_0=0.073$ (no-rot) to $\beta_\mathrm{s}=0.207$ 
whether $^{152}$Gd is situated within the applicability of the RPA. 
In order to look into this, we compare the interband/in-band ratio of $B(E2)$ 
to the case of the wobbling that is another example of strong interband $E2$ transitions 
previously accounted for in terms of the RPA. In the present case, 
the calculated ratio of the $j=1$ (interband) and $j=2$ (in-band) terms, 
\begin{equation*}
\frac{B(E2,0^+_2\rightarrow2^+_1)}{B(E2,0^+_2\rightarrow2^+_2)}
=\frac{B(E2,0^+_2\rightarrow2^+_1)}{B(E2,0^+_1\rightarrow2^+_1)}
=\frac{B(E2,0^+_2\rightarrow2^+_1)}{5\times B(E2,2^+_1\rightarrow0^+_1)},
\end{equation*}
amounts to 0.12 (no-rot) and 0.52 (GIR). 
The wobbling excitations in the triaxially super/strongly deformed states in Lu isotopes were 
observed~\cite{OHJ,SHH,AMH}, calculated in terms of the RPA~\cite{MSM,AND,SSM,SS,FD}, 
and in other models~\cite{Ha,CMZ,TT}. Their ratios are 
$\frac{B(E2,I\rightarrow I-1)}{B(E2,I\rightarrow I-2)}\sim0.2$. 
The associated fluctuation, the wobbling angle $\theta$, is about 0.44 radian, 
for example, which fulfills a criterion of validity of the small amplitude approximation, 
$\tan{\theta}\simeq\theta$~\cite{SS}. In comparison of the present ratio, 0.12 that 
is directly given by the RPA, with that of the wobbling case, 0.2, we consider that 
the RPA is applicable to the $\beta$ vibration in $^{152}$Gd. Then, the effective value, 
$\beta_0=0.154$ (GIR), looks to indicate that, even if there exists some difference between 
the equilibrium deformations of $0^+_1$ and $0^+_2$ that is ignored in the present model, 
it would be of little relevance as conjectured in Ref.~\cite{DH}.

This strong interband transition is an outcome of strong ground-state correlations, 
in other words, large backward amplitudes. These backward amplitudes stem from 
time-reversal pairs near the Fermi surface, such as $|\phi|=$ 0.916, 0.959 and 0.699 
for $(\nu[532]\frac{3}{2})^2$, $(\nu[530]\frac{1}{2})^2$ and $(\nu[521]\frac{3}{2})^2$, 
respectively, in the present case as discussed in the case of $^{154}$Gd below. 
Consequently, the pair transfer 
crosssection is also expected to be enhanced without recourse to the shape 
coexistence. Actually it is shown in Ref.~\cite{MWC2} that the 
$^{154}$Gd $(p,t)$ $^{152}$Gd crosssection is stronger for $0^+_2$ than for 
$0^+_1$. This fact does not contradict the RPA result. 

(2) $^{154}$Gd

The $0^+_2$ state at $E=681$ keV in this nucleus is another candidate of typical 
$\beta$ vibrations~\cite{Ga}. In contrast, the $0^+_3$ state at $E=1182$ keV is 
thought to have a smaller deformation~\cite{SDN} and to be a pairing 
isomer~\cite{RB,KWK1}. Higher-lying states above 1 MeV were also 
investigated~\cite{KWK2}. Although a possibility of interpreting the $2^+$ state 
at $E=1531$ keV as the $\beta\otimes\gamma$ double excitation assuming that the 
$0^+_2$ is a $\beta$ vibration is reserved, the authors of this reference suggest that the 
$0^+_2$ state has a shape different from that of the ground state rather than 
is the $\beta$ vibration on top of it based on the non-existence of the two-phonon 
$\beta$ vibrational state. A similar argument was done also for 
$^{152}$Sm~\cite{KWG,GKW}, but the non-existence of the two-phonon $\beta$ 
vibrational state does not necessarily mean that of the one phonon. 

Reference~\cite{SS1} further proceeds in this direction; the $0^+_2$ is also a 
pairing isomer with a smaller deformation although transition properties are 
not considered. Microscopically the main component of the $0^+_2$ is 
$(\nu[505]\frac{11}{2})^2$ in this scenario. An important consequence of their argument 
is that this scenario leads to the non-existence of the $0^+_2\otimes\nu[505]\frac{11}{2}$ 
band in the adjacent odd nuclei, $^{153}$Gd and $^{155}$Gd, because of the blocking 
effect. In an accompanying paper~\cite{SS2}, the authors studied $^{155}$Gd and 
concluded that the $0^+_2\otimes\nu[521]\frac{3}{2}$ and the $\gamma\otimes\nu[505]\frac{11}{2}$ 
bands exist but the $0^+_2\otimes\nu[505]\frac{11}{2}$ does not. Note here that 
the $\gamma\otimes\nu[505]\frac{11}{2}$ is an unusually high-$K$ band. That was also observed 
already in Ref.~\cite{HOH}; the spin assignments of these two works differ by one 
unit from each other. 

\begin{table}[htbp]
 \caption {The results of the RPA calculation for each configuration. Among them, 
$\Delta_\mathrm{n}$ and $\hbar\omega_\beta=E_{0^+_2}$ of $^{154}$Gd are fitted to the data 
(TABLE~\ref{table1}). $\sum\phi^2$ denotes the sum of the squared backward 
amplitudes in the RPA phonon. $t_{Q_0^{(+)}}^2$ is equal to $B(E2,0^+_2\rightarrow2^+_1)$ 
without the rotational effect in the case of the even-even nucleus. }
 \label{table2}
\begin{center}
\begin{tabular}{cccccc} \hline\hline
nucleus & band & $\Delta_\mathrm{n}$ (MeV) & $\hbar\omega_\beta$ (MeV) & $\sum\phi^2\,$ & $t_{Q_0^{(+)}}^2$ (W.u.) \\ 
\hline
$^{154}$Gd & ground              & 1.28 & 0.681 & 3.07 & 22.7 \\
$^{155}$Gd & $[521]\frac{3}{2} $ & 1.10 & 0.974 & 1.39 & 13.6 \\
$^{155}$Gd & $[505]\frac{11}{2}$ & 1.16 & 1.475 & 0.47 &  8.9 \\
\hline\hline
\end{tabular}
\end{center}
\end{table}

Here we examine the results of RPA calculations on 
1) the ground state of $^{154}$Gd, 
2) the $[521]\frac{3}{2}$ state of $^{155}$Gd, and 
3) the $[505]\frac{11}{2}$ state of $^{155}$Gd, in order to see how the $\beta$ vibrational 
calculation can account for the observed properties. Calculations for the odd-$A$ 
cases are done on their ground states specified by blocking an appropriate 
quasiparticle state~\cite{SM2} obtained by the calculation for $^{154}$Gd. 
The difference between $^{153}$Gd and $^{155}$Gd 
is specified by the chemical potential that gives the correct particle number. 
The interaction strengths $G_\mathrm{n}$, $G_\mathrm{p}$, and $\kappa_K^{(+)}$ are kept 
unchanged. The results are summarized in TABLE~\ref{table2}. 

In the phonon wave function of the $^{154}$Gd case, 1), large backward amplitudes 
$|\phi|$ stem from time-reversal pairs of prolate (low-$\Omega$) orbitals, such as 
$|\phi|=$ 0.771, 0.411 and 0.631 for $(\nu[660]\frac{1}{2})^2$, $(\nu[521]\frac{3}{2})^2$ 
and $(\nu[651]\frac{3}{2})^2$, respectively. 
In contrast, the only large forward amplitude is that of $(\nu[505]\frac{11}{2})^2$, 
$|\psi|=0.995$. This proves that the main origin of the collectivity is different 
from the main forward component. In the case of the $[521]\frac{3}{2}$ of $^{155}$Gd, 2), 
collectivity is reduced by blocking one of prolate orbitals but the resulting 
$\beta$ vibration is still collective enough. In the case of the $[505]\frac{11}{2}$ of 
$^{155}$Gd, 3), the wave function is changed dramatically by loosing the main 
forward component. Consequently the $K=0$ strength is pushed up to higher energies 
but still distinguishable from other non-collective states. We also confirmed that 
the $\gamma$ vibration is almost not affected because there are no neutron 
quasiparticle states that constitute $K=2$ pairs with $[505]\frac{11}{2}$. 
These results prove that the characteristics of the spectra of $^{154}$Gd and 
$^{155}$Gd can be accounted for in terms of the $\beta$ vibration. However, 
it should be noted that the isomerism of $0^+_3$ and the difference in the 
crosssections of $(p,t)$ and $(t,p)$ transfers to $0^+_2$ are out of the scope of 
the present calculation that does not contain the quadrupole pairing. 

(3) $^{156}$Gd

\begin{figure}[htbp]
 \includegraphics[width=7cm]{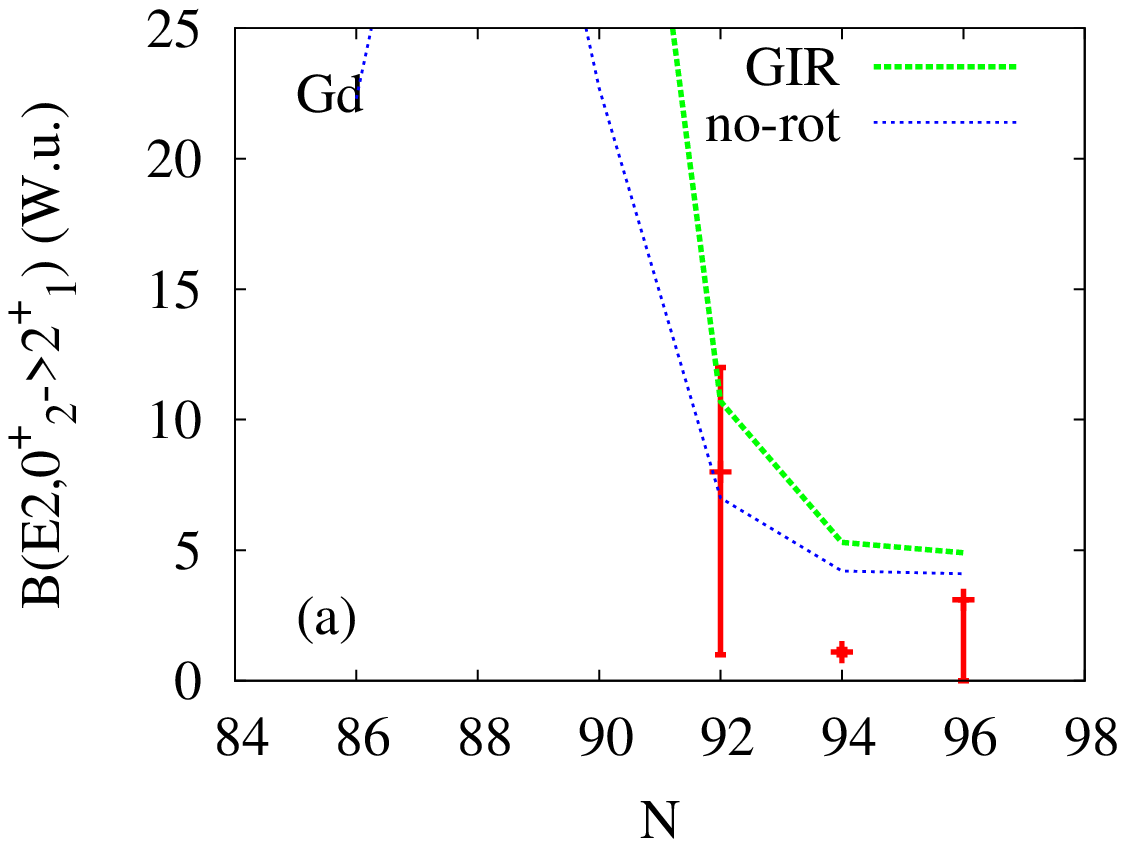}
 \includegraphics[width=7cm]{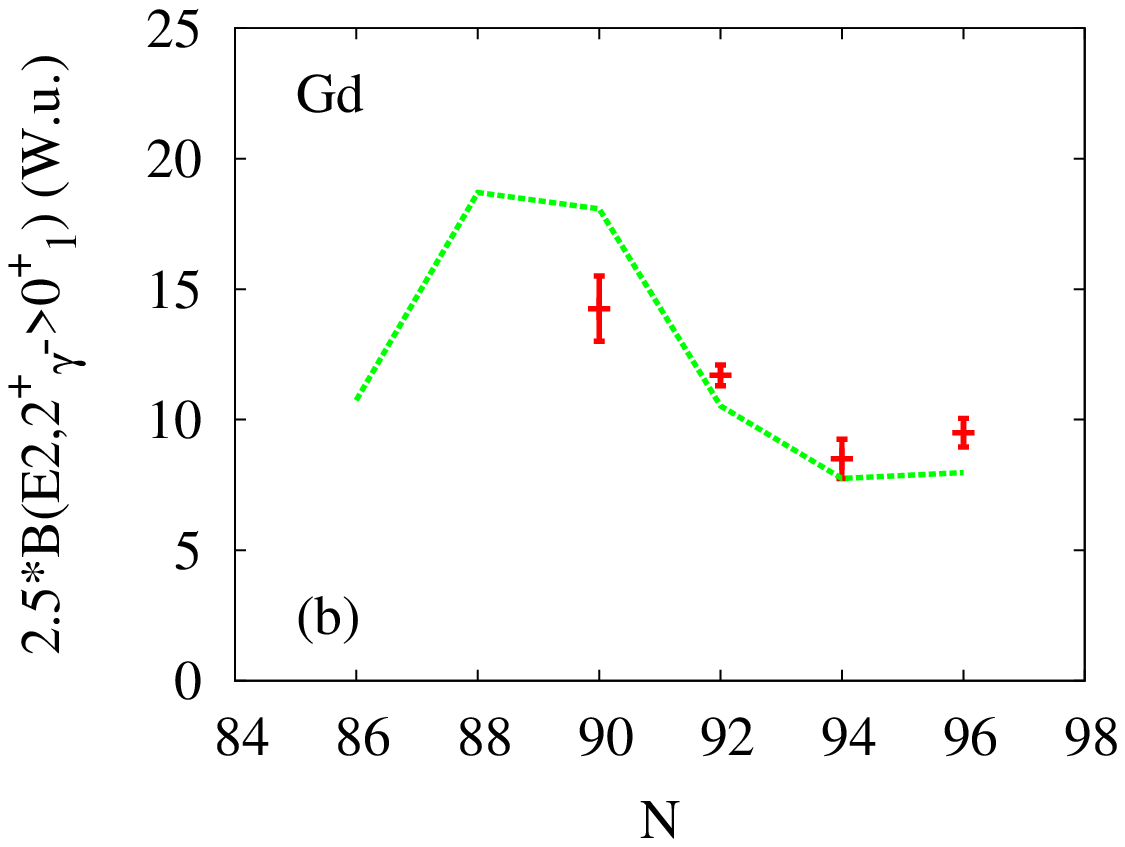}
 \caption{(Color online) (a) Low-$B(E2)$ part of Fig.~\ref{fig1}(a). 
(b) $B(E2,2^+_\gamma\rightarrow0^+_1)$ multiplied by $\frac{5}{2}$ in order to 
compare the matrix elements with (a). The rotational effect does not appear 
in the latter. Data are taken from Ref.~\cite{IAEA}.}
 \label{fig5}
\end{figure}

Figure~\ref{fig5} compares $B(E2)$ of (a) $\beta$ and (b) $\gamma$ vibrations of 
heavier isotopes. This indicates that $^{156}$Gd is located at the point where 
the $\beta$ and $\gamma$ vibrations have similar transition matrix elements 
as well as excitation energies (TABLE~\ref{table1}). In this sense, $^{156}$Gd 
can be regarded as one of good examples of Bohr-Mottelson's picture of deformed 
nuclei. 

(4) $^{158}$Gd

Figure~\ref{fig5}(a) also indicates that the observed $B(E2,0^+_2\rightarrow2^+_1)$ 
in $^{158}$Gd looks evidently smaller than expected from the systematics. Actually this 
is one of the curious properties that have this nucleus extensively studied but have 
not been resolved yet. Since an early study~\cite{GRB}, the $0^+_3$ state at 
$E=1452$ keV has been known to be more collective than the $0^+_2$ state at $E=1196$ keV. 
Although $\rho^2(E0,0^+_2\rightarrow0^+_1)$ has not been reported up to now, 
$\rho^2(E0,2^+_2\rightarrow2^+_1)=(0.72\pm0.21)\times10^{-3}$ and 
$\rho^2(E0,2^+_3\rightarrow2^+_1)=(25\pm4)\times10^{-3}$ were reported in that work and 
reevaluated as $\le0.8\times10^{-3}$ and $(17\pm3)\times10^{-3}$, respectively, in 
Ref.~\cite{WZC} for the rotational-band members. The quadrupole transition strengths 
were measured much later in Ref.~\cite{BJZ} as $B(E2,0^+_2\rightarrow2^+_1)=1.1$ W.u. and 
$B(E2,0^+_3\rightarrow2^+_1)=2.1$ W.u., see also Ref.~\cite{LOA}. In addition to the fact 
that the latter is larger, both of them are smaller than expected for ordinary 
$\beta$ vibrations. Later a large number of $0^+$ states were reported~\cite{LAT}. 
Moreover, in Ref.~\cite{LOA}, 
$B(E2,0^+_n\rightarrow2^+_1)$ were measured for a lot of states up to $0^+_{10}$. 
This result proves that the $E2$ strengths are strongly fragmented and pushed up to 
higher energies; the largest one is $B(E2,0^+_8\rightarrow2^+_1)=7.7^{+1.5}_{-0.7}$ W.u. 
References~\cite{ZZC,NRR} suggest a contribution of two-phonon octupole vibration 
to producing large number of $0^+$ states based on the geometrical collective model 
and the IBA model. In the projected shell model~\cite{SAZ}, the excited energies and 
the number of $0^+$ states are accounted for by two- and four-quasiparticle states but 
associated $B(E2)$ are much smaller than observed. Lo Iudice et al.~\cite{LSS} and 
Ger\c{c}eklioglu~\cite{Ge} performed RPA calculations. The former includes the 
quadrupole pairing interaction. The resulting number of $0^+$ states is less than 
observed in the RPA calculation but quasiparticle-phonon couplings with octupole two-phonon 
states improve the result. The latter includes the spin-quadrupole interaction. 
The number of $0^+$ states is reproduced without an octupole-octupole interaction. 
Both calculations, however, failed to account for the character of the $0^+_2$ and $0^+_3$ states. 
\begin{figure}[htbp]
 \includegraphics[width=7cm]{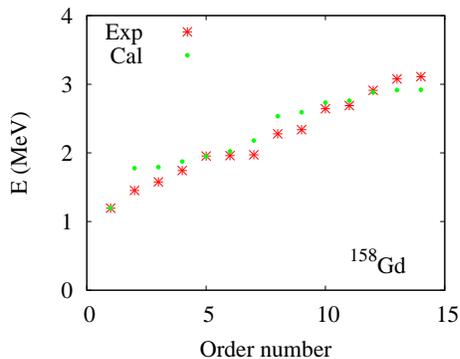}
 \caption{(Color online) Experimental and calculated distribution of excited $0^+$ states 
in $^{158}$Gd. Data are taken from Ref.~\cite{LAT}.}
 \label{fig6}
\end{figure}
Our RPA result for the distribution of excited $0^+$ states is presented in Fig.~\ref{fig6}. 
The energy of the lowest excitation is fitted by adjusting the interaction strength 
$\kappa_0^{(+)}$. This figure shows that the overall distribution is reproduced quite well 
without an octupole-octupole interaction but the obtained $0^+_3$ is not collective. 
None of higher states have $B(E2)$ strengths larger than 1 W.u. 

One of possible origins of quadrupole collectivity at high energies conjectured in 
Ref.~\cite{LOA} is the two-phonon $\gamma$ vibration. The $K=0$ two-phonon $\gamma$ vibration 
is known only in $^{166}$Er~\cite{GKL} although the $K=4$ ones are known more as briefly 
reviewed in Ref.~\cite{MM2G}. The quasiparticle-phonon coupling model calculation in 
Ref.~\cite{LSS} looks to include such a type of excitation but reported $B(E2)$ are much 
smaller. 

(5) $^{160}$Gd

Very recently an upper limit of $B(E2,0^+_2\rightarrow2^+_1)$ was reported~\cite{LCA}. 
The calculated value is slightly larger than the reported upper limit as shown in 
Fig.~\ref{fig5}(a). But it is open whether there is a problem similar to $^{158}$Gd. 

\section{Conclusions}

The long-debating problem of the characterization of the $0^+_2$ states in Gd isotopes 
has been revisited. The model adopted is a traditional mean field plus RPA. The 
doubly-stretched quadrupole-quadrupole interaction is used. The rotational effect on 
the transition strengths are accounted for by that on the intrinsic matrix elements 
based on the generalized intensity relation. Calculations have been done paying attention 
to properties of rotational bands. 

The most decisive property to characterize the $0^+_2$ states is $B(E2,0^+_2\rightarrow2^+_1)$. 
Its steep $N$ dependence ranging two orders of magnitude is nicely reproduced. 
In particular, those in lighter isotopes, $^{152}$Gd and $^{154}$Gd have been shown 
to be understandable as $\beta$-vibrational excitations on top of deformed ground states 
as early thought~\cite{Ga}. To this end, an implication of the strengths of $B(E2)$ 
and rotational properties for the former, and the relation to the spectra of the 
adjacent odd-$A$ nucleus for the latter have been investigated. Consequently the present 
calculation supports the picture of Ref.~\cite{Bu}. The monopole transition strength, 
$\rho^2(E2,0^+_2\rightarrow0^+_1)$, is also thought to be sensitive to the shape 
deformation/coexistence. The available data have been reproduced fairly well within 
the present model but data are still too scarce to utilize for discriminating different 
theoretical pictures. 

Looking at relatively weak $B(E2)$ in heavier isotopes more closely, however, a disagreement 
remains in $^{158}$Gd; a strong fragmentation of $B(E2)$ strengths to higher energies is not 
accounted for in the present model as well as preceding works.

\end{document}